\documentclass[twoside,twocolumn]{article}
\usepackage[super,sort&compress,comma]{natbib} 
\usepackage[version=3]{mhchem}
\usepackage{times}
\usepackage{balance} 

\usepackage[dvips]{graphicx} 
\usepackage{epsfig}
\usepackage{lastpage}
\usepackage[format=plain,justification=raggedright,singlelinecheck=false,font=footnotesize,labelfont=bf,labelsep=space]{caption} 
\usepackage{fancyhdr}
\usepackage[]{subfigure} 

\usepackage[usenames]{color}

\begin{document}

\thispagestyle{plain}
\fancypagestyle{plain}{
\renewcommand{\headrulewidth}{1pt}}
\renewcommand{\thefootnote}{\fnsymbol{footnote}}
\renewcommand\footnoterule{\vspace*{1pt}%
\hrule width 3.4in height 0.4pt \vspace*{5pt}} 
\setcounter{secnumdepth}{5}

\makeatletter 
\def\subsubsection{\@startsection{subsubsection}{3}{10pt}{-1.25ex plus -1ex minus -.1ex}{0ex plus 0ex}{\normalsize\bf}} 
\def\paragraph{\@startsection{paragraph}{4}{10pt}{-1.25ex plus -1ex minus -.1ex}{0ex plus 0ex}{\normalsize\textit}} 
\renewcommand\@biblabel[1]{#1}            
\renewcommand\@makefntext[1]%
{\noindent\makebox[0pt][r]{\@thefnmark\,}#1}
\makeatother 
\renewcommand{\figurename}{\small{Fig.}~}

\setlength{\arrayrulewidth}{1pt}
\setlength{\columnsep}{6.5mm}
\setlength\bibsep{1pt}

\twocolumn[
  \begin{@twocolumnfalse}
\begin{center}
\noindent\LARGE{\textbf{Charge and Synthesis Modification in PMMA-PDMS Copolymer Model Colloids}}
\vspace{0.6cm}

\noindent\large{\textbf{Rodrigo S\'anchez\textit{$^{a}$}
}}\vspace{0.5cm}
\end{center}

\noindent \normalsize{
Colloidal charge in non-polar media is less well understood than its aqueous counterpart and due to van der Waals forces its study is particularly challenging in highly confined systems, for which on the other hand large long-range electrostatic repulsions are heavily screened in aqueous media. Evidence from videomicroscopy is presented that a model system governed with surfactant-mediated charging is governed by the steric stabilizer's specific chemistry. 
}
\vspace{0.6cm}
 \end{@twocolumnfalse}
]

\section{Introduction}



\footnotetext{\textit{$^{a}$~Departamento de F\'isica, Universidad Veracruzana, Zona Universitaria, Xalapa 91000, Veracruz, Mexico. Email: rodrsanchez@uv.mx}}

Colloidal charge and electrostatic repulsion in non-polar solvents are of considerable scientific and industrial interest\cite{868, Comiskey1998, 3958}, despite their typically modest charges\cite{868}. 

Charging has been demonstrated in sterically stabilized poly(methyl methacrylate) (PMMA) colloids in non-polar solvents before\cite{56, 4438, Roberts2007a, Roberts2008, Sainis2008, Kemp2010, Sanchez2011}, and there is evidence that, in the case of surfactant-mediated charging, it is a result of surfactant adsorption \cite{Roberts2008, Kemp2010} ocurring, at least for systems sterically stabilized by poly(12-hydroxystearic acid) (PHSA), which has also been used as a stabilizer in the synthesis of PMMA colloids with polar surface functionality\cite{Lee1994}, and involving the surfactant sodium di-2-ethylhexylsulfosuccinate (AOT), thought to be adsorbed at the core-stabilizer interface\cite{Kemp2010}. AOT-mediated charging has been shown to be sensitive to surface hydrophilicity in other systems \cite{4908}. Significant charging, mediated by the surfactants sorbitan trioleate and monooleate (span 80) in hydrocarbons, of PMMA colloids with a different surface chemistry has also been observed\cite{Espinosa2010}.

When PMMA colloids have been synthesized in-house, this has typically been done using methods 
based on the dispersion polymerization procedures of Antl \emph{et al}.\cite{177}; in such preparations, the PHSA stabilizer covalently binds to methacrylic acid (MAA) groups in the poly(MMA-\emph{co}-MAA) core (mostly comprised of MMA groups), for which there is evidence that it contains caged PHSA\cite{Pekcan1983, Kemp2010}. However, PHSA synthesis is relatively inconvenient, and sterically stabilized PMMA colloidal spheres have been synthesized using commercially available high molecular weight poly(dimethyl siloxane) (PDMS) instead\cite{Keville1991, Mohraz2005}. In the present work, a system of poly(MMA-\emph{co}-MAA) colloidal spheres, sterically stabilized by PDMS-\emph{g}-PMMA copolymer chains, was studied in a quasi-2D cell. In such cells, hard sphere behavior has been observed for charged colloids in water even in the absence of added electrolytes\cite{Santana-Solano2001, Santana-Solano2005}, which is likely to arise from heavy screening due to countercharges from the confining glass walls. In contrast, it is plausible that for charged systems in non-polar solvents the screening would be less and thus deviations from hard sphere behavior could be observed. This idea was tested in the present work for surfactant-mediated charging.

\section{Materials and Methods}

The monomers methyl methacrylate and methacrylic acid (both from Sigma-Aldrich) were used, with benzoyl peroxide as initiator, and trimethylsiloxy terminated, 300, 000 cSt poly(dimethyl siloxane) (PDMS) (Gelest) was copolymerized with MMA to form the steric stabilizer. All materials were used as received save for the MMA, which was passed through an inhibitor removal column (Sigma-Aldrich). The protocol of Mohraz and Solomon\cite{Mohraz2005} was used, without a dye, modified by the incorporation of MAA (3 wt \% of monomers) and minor modifications such as the use, for experimental convenience, of an equivalent volume of cyclohexane instead of hexane as dispersant, resulting in 0.687 $\pm$ 0.023 $\mu$m particles. Larger PMMA spheres (0.902 $\pm$ 0.060 $\mu$m) were synthesized by increasing the monomer concentration wihout MAA, and were used as spacers. Particle sizes were determined by dynamic light scattering. 

The solvents decalin (mixture of \emph{cis} and \emph{trans}) and dodecane (Sigma-Aldrich) and the surfactants span 80 and AOT (Sigma-Aldrich) were used as received. For both surfactants, concentrations of 5 mM were used, in both dodecane and decalin in the case of AOT, and in decalin only in the case of span 80. The particles were transferred into the surfactant solutions by repeated centrifugation, removal of the supernatant and re-dispersion.

Samples and cells were prepared, and optical microscopy was carried out, according to the protocols of Arauz-Lara \emph{et al}.\cite{Santana-Solano2001, Santana-Solano2005}. Where applicable, clean, dry coverslips and microscope slides were hydrophobed by immersion in an octadecyltrichlorosilane solution in cyclohexane followed by sonication in chloroform.

\section{Results and Discussion}

\begin{table}[h]
\small 
\caption{\ The various systems examined, and the key observations made.} 
\label{t1}
\begin{tabular*}{0.5\textwidth}{@{\extracolsep{\fill}}lccc}
  \hline Solvent & Glass & Surfactant & Observations \\
  \hline
  Any & Hydrophilic & Any & Coagulation \\
  Decalin & Hydrophobic & AOT & No visible repulsion \\
  Decalin & Hydrophobic & Span 80 & No visible repulsion \\  
  Dodecane & Hydrophobic & AOT & Coagulation \\

  \hline
\end{tabular*}

\end{table}

As shown in table \ref{t1}, coagulation was succesfully prevented in systems with hydrophobic glass by the use of decalin, whose refractive index is similar to that of the particles, while the use of hydrphylic glass and / or dodecane resulted in coagulation. Qualitative inspection of the videomicroscopy results revealed no noticeable difference between the systems with surfactant and hard sphere systems; specifically, no evidence was observed of any hindrance of close interparticle approach, nor any evidence of Yukawa crystallization. 

The existence of very modest net charges, and of feeble interparticle electrostatic interactions, cannot be ruled out without further experimental measurements, for example by using single particle optical electophoresis (SPOM) or by the detailed characterization of the dynamics and/or the structure of the system. Nevertheless, the existence of a non-fluid phase in the samples examined can be reasonably ruled out. Given the modest concentrations of surfactant used, and hence the large screening lengths expected, it is unlikely that this is merely the result of screening, and is more likely the result of either modest overall surfactant adsorption, or of any such adsorption being non-preferential in terms of charge. Either effect would strongly suggest effects associated with the steric stabilizer's specific chemistry, an effect not specifically considered in previous work regarding the charging effects of AOT adsorption onto PMMA colloids\cite{4437, Roberts2007a, Roberts2008, Kemp2010, Sanchez2011}. It is an open question whether these affects are associated with the covalent bonds between the MAA and PHSA groups that are absent in the particles used for the present work or, as seems more likely to this author, with the polar groups in the PHSA, whether that covalently bound to the surface or that caged within the core. In particular, the observation that the addition of modest amounts of AOT to PHSA-stabilized PMMA colloids synthesis \cite{Sanchez2010} results in coagulation suggests that AOT adsorption is not prevented by the absence of a PHSA-MAA covalent bond, suggesting in turn that charging in such systems is due to PHSA polar groups.


\section{Conclusions}

Surfactant-mediated charge in these model systems appears to be sensitive not only to the specfic chemistry of the colloid surface itself, but to that of the steric stabilizer and/or the bonds between said stabilizer and the core. It is possible polar groups in the stabilizer itself are required for the system to exhibit charge leading to long-range repulsion, as that is a key difference between the PHSA and PDMS-PMMA stabilizers. 

\subsection*{Acknowledgements}

It is a pleasure to acknowledge the support of JL Arauz-Lara, without whose generosity the present work would not have been possible.

\footnotesize{

\bibliography{bib}

\bibliographystyle{rsc}

}
\end{document}